\documentstyle[11pt,times,psfig]{article}

\begin{document}
\noindent{
{\Large\bf Fermilab-Conf-98/379 (1998)\\
LANL Report LA-UR-98-5716 (1998)\\
}
}
Talk given at the {\em Fourth Workshop on Simulating Accelerator Radiation
Environments (SARE4)}, \\
Knoxville, Tennessee, September 14-16, 1998

\vspace{1cm}
\begin{center}
{\Large\bf MARS Code Developments\renewcommand{\thefootnote}{*}\footnote{
Work supported by the Universities Research Association, Inc., under contract
DE-AC02-76CH00300 with the U.~S.~Department of Energy.}\\}

\vspace{0.5cm}
N.~V.~Mokhov, S.~I.~Striganov$^{\dagger}$ and A.~Van Ginneken \\
{\footnotesize \em Fermi National Accelerator Laboratory, 
P.O.~Box 500, Batavia, IL 60510}\\
{\footnotesize \em $^{\dagger}$ Now at Institute for High-Energy Physics, 
Protvino, Moscow Region, Russia}\\

\vspace{0.3cm}
S.~G.~Mashnik and A.~J.~Sierk \\
{\footnotesize \em Los Alamos National Laboratory, T-2, MS-B243, 
Los Alamos, NM 87545}\\
\vspace{0.3cm}
J. Ranft\\
{\footnotesize \em University of Siegen, Siegen, D-57068, Germany}
\end{center}

\begin{abstract}
Recent developments in the physical model of 1~MeV to 100~TeV hadron and lepton 
interactions with nuclei and atoms are described. 
These include a new nuclear cross section library, a model for soft pion 
production, the cascade-exciton model, the dual parton model, 
deuteron-nucleus and neutrino-nucleus interaction models, detailed description 
of $\mu^-$, $\pi^-$ and $\overline p$ absorption and a unified treatment of 
muon and charged hadron electromagnetic interactions with matter. 
New algorithms are implemented into the \textsc{mars13(98)} Monte Carlo code 
and benchmarked against experimental data. 
The code capabilities to simulate cascades and generate a variety of results 
in complex media have been also enhanced. 
\end{abstract}

\section{Introduction}

The \textsc{mars} Monte Carlo code system, being developed over 24~years,
allows fast and reliable inclusive simulation of three-dimensional hadronic and 
electromagnetic cascades in shielding, accelerator and detector components in
the energy range from a fraction of an electron-volt up to about 
100~TeV~\cite{mars-web}.
The reliable performance of the \textsc{mars13} code~\cite{mars95,mars97} has 
been demonstrated in numerous applications at Fermilab, CERN, KEK and other 
centers as well as in special benchmarking 
studies~\cite{nakane,hideo,KEK12gev}. 
Recently, challenging applications at Fermilab have induced further 
significant developments of the code's physical model and its scoring 
capabilities. 
New developments in both the electromagnetic and strong interaction sectors 
as well as other enhancements to the current version 
-- \textsc{mars13(98)}~\cite{mars-web} -- are briefly described in this paper.

\section{Nuclear Cross Sections}

\hspace{6mm}
\textbf{Hadron-nucleon cross sections.} New compilations and parameterizations 
of elastic and inelastic $\sigma_{hN}$ are implemented covering a hadron 
kinetic energy range 1~MeV$<$E$<$100~TeV. 
Total cross sections, $\sigma_{tot}$, from 1~MeV to 10~GeV for $p, n, \pi^+$ 
and $\pi^-$ are as predicted by
the new improved algorithm~\cite{CEM98} of the Cascade-Exciton Model (CEM)~\cite{CEM}
code \textsc{cem95}~\cite{CEM95} while for $K^+$, $K^-$ and $\overline p$ data 
compilations are used~\cite{RPP98}. 
Parameterizations from~\cite{RPP98} are relied upon for all particles between
10~GeV and 100~TeV.
Elastic cross sections, $\sigma_{el}$, from 10~MeV to 10~GeV for $p, n, \pi^+$ 
and $\pi^-$ are likewise from~\cite{CEM98} with interpolation of 
data~\cite{RPP98} for $K^+$, $K^-$ and $\overline p$. 
Parameterizations from~\cite{RPP94} are used between 10 and 200~GeV.
For energies 200~GeV$<$E$<$100~TeV, the optical theorem with `universal 
slope'~\cite{Burq83} is applied. 
Fig.~1(a) shows comparison of data and \textsc{mars} results on $\sigma_{tot}$ 
and $\sigma_{el}$ for $\pi^{-}p$ collisions.

\begin{sloppypar}
\textbf{Hadron-nucleus cross sections.} New compilations, parameterizations
and integration algorithms for total, inelastic, production and elastic 
$\sigma_{hA}$ are introduced into the code.
Total, inelastic and elastic cross sections from 1~MeV to 5~GeV are described
using new compilations and improved interpolation 
algorithms~\cite{Bar93,crosec}.
At higher energies (5~GeV$<$E$<$100~TeV), $\sigma_{tot}$, $\sigma_{in}$, 
$\sigma_{prod}$ and $\sigma_{el}$ are calculated in the framework of the 
Glauber multiple scattering theory with the above $\sigma_{hN}$ as an input.
The nucleon density distribution in nuclei is represented as the 
symmetrized Fermi function with the parameters of~\cite{Alk78} for medium and 
heavy nuclei ($Z>10$) and the ones of~\cite{Bur76} for $Z<10$. 
An example is shown in Fig.~1(b) for neutron-nucleus $\sigma_{tot}$ as 
calculated with this algorithm (solid line) and with 
the improved algorithm~\cite{crosec} (dashed line).
\end{sloppypar}

\textbf{Photon-nucleus cross sections.}
Data compilation and interpolation algorithm for $\sigma_{\gamma N}$ with
phenomenological $A$-dependence for $\sigma_{\gamma A}$ are as described 
in~\cite{mokstr96}.

\begin{figure}[b!]
\begin{minipage}[b!]{.52\linewidth} 
\psfig{figure=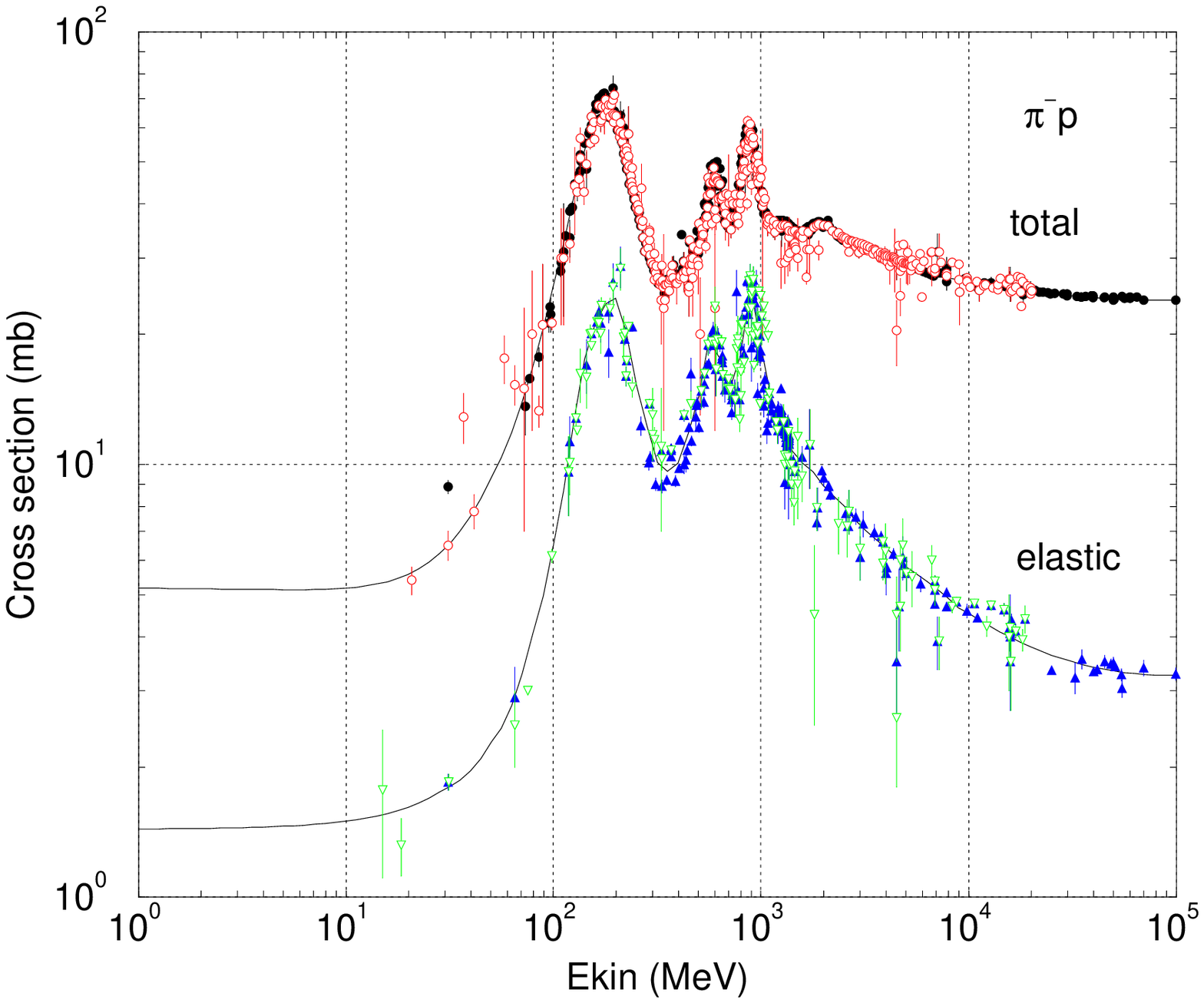,width=\linewidth,angle=0}
\end{minipage}\hfill
\begin{minipage}[b!]{.41\linewidth} 
\psfig{figure=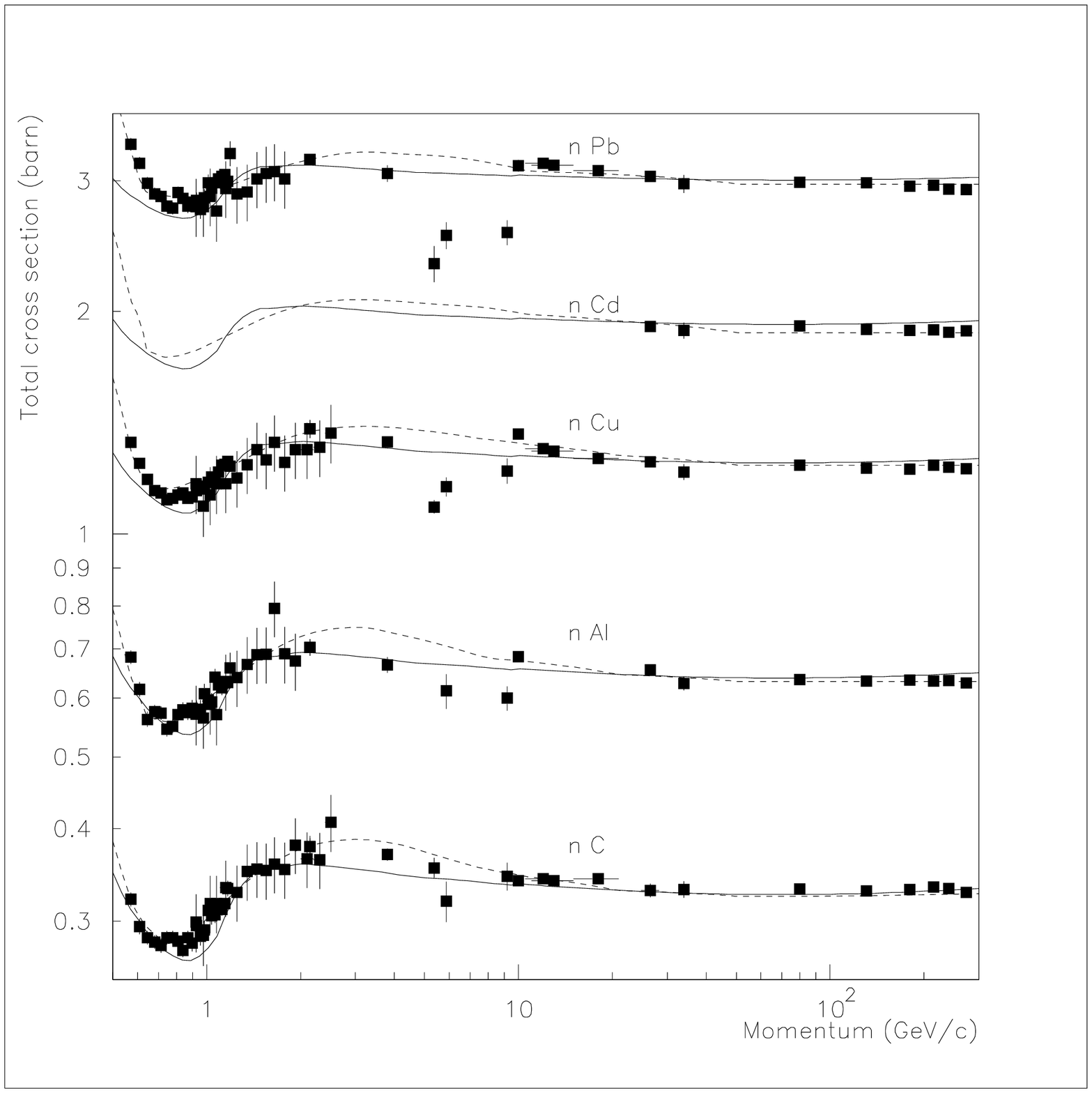,width=\linewidth,angle=0}
\end{minipage}

\vspace*{0.3cm}
Figure 1.
{\sc mars} cross sections in comparison with experimental data:
(a) $\sigma_{tot}$ and $\sigma_{el}$ for $\pi^{-}p$ collisions as a function 
of pion kinetic energy;
(b) $\sigma_{tot}$ for neutrons {\em vs} beam momentum.
\end{figure}

\section{Nuclear Reactions from 10 MeV to 100 TeV}

\begin{sloppypar}
\hspace{6mm}
\textbf{Improved description of hadron-nucleus elastic scattering.}
Two algorithms are implemented into \textsc{mars} to better describe 
hadron-nucleus elastic scattering for 10~MeV$<$E$<$5~GeV: one from the code
\mbox{\textsc{lahet}}~\cite{lahet} while the other is based on a phenomenological 
formula~\cite{sychev}. 
As one can see from Fig.~2(a), more work is still needed here.
However, at higher energies, the model used in \textsc{mars} for both 
coherent and incoherent components of $d\sigma /dt$ is quite consistent with
experiment.
\end{sloppypar}

\begin{figure}[t!]
\begin{minipage}[t!]{.52\linewidth} 
\psfig{figure=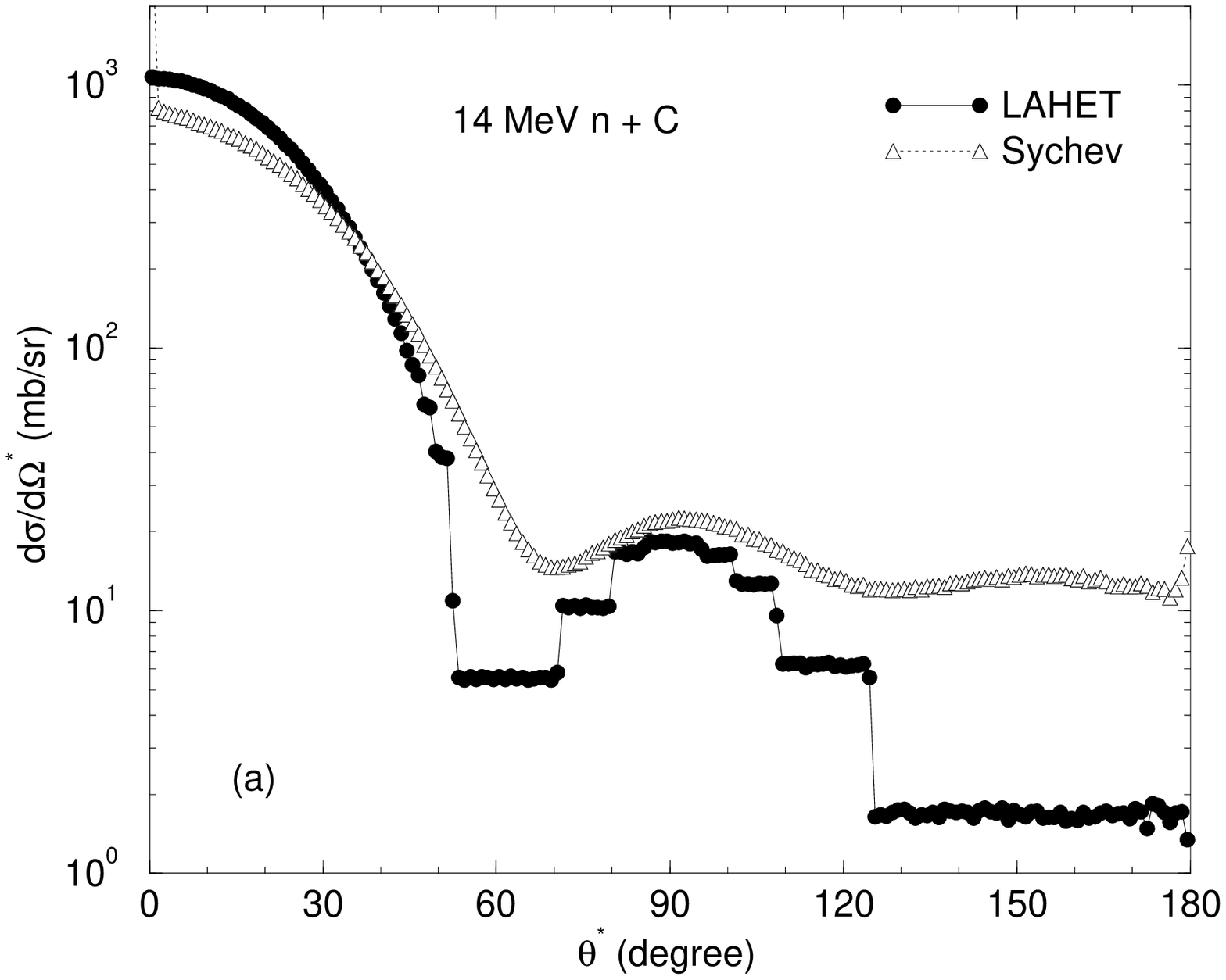,width=\linewidth,angle=0}
\end{minipage}\hfill
\begin{minipage}[t!]{.41\linewidth} 
\psfig{figure=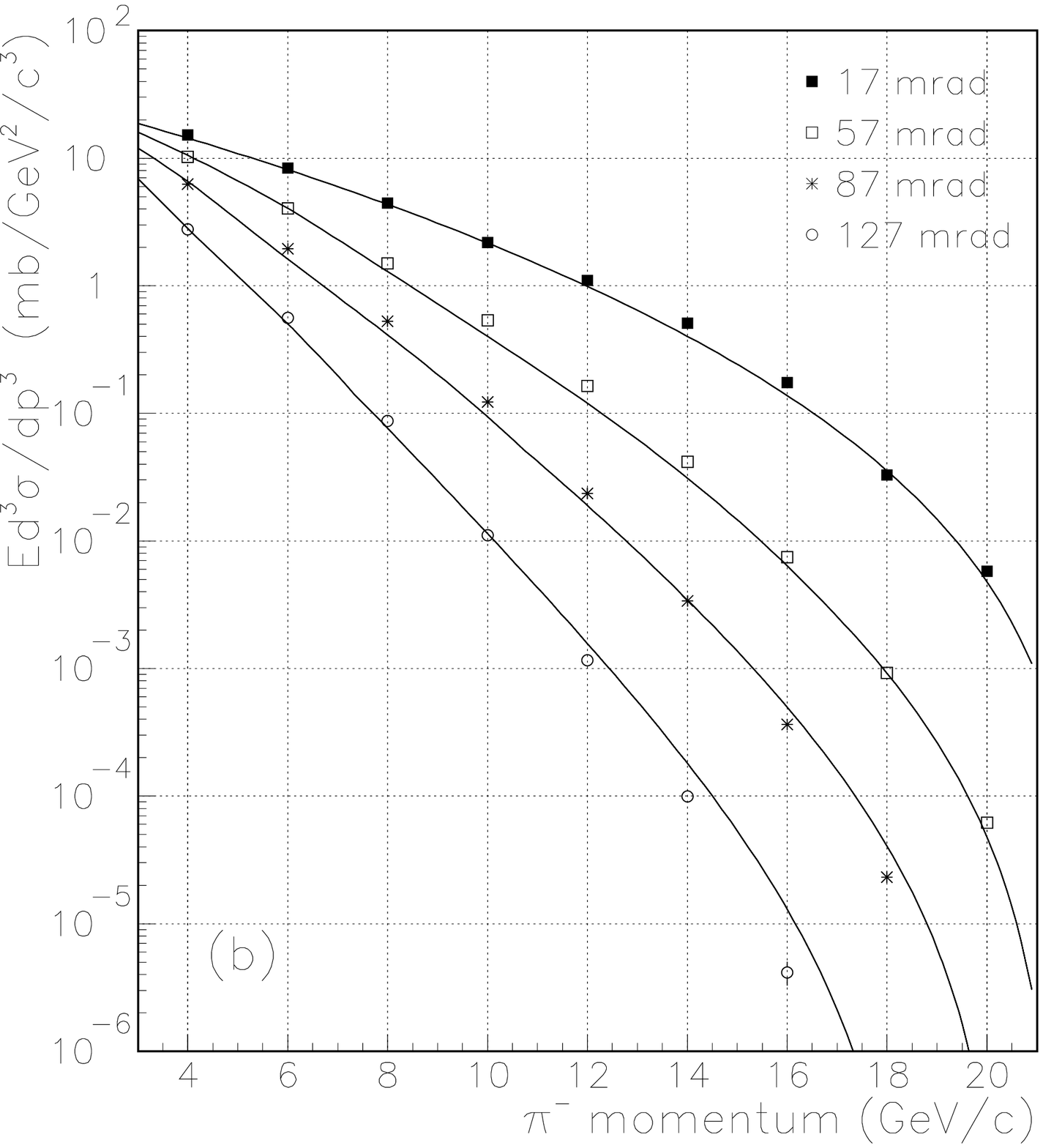,width=\linewidth,angle=0}
\end{minipage}

\vspace*{0.3cm}
Figure 2.
(a) Angular distribution of 14~MeV neutrons in center-of-mass, 
scattered off carbon as calculated according to~\textsc{lahet}~\cite{lahet} 
and \cite{sychev};
(b) Pion spectra for $pp \rightarrow \pi^-X$ at four angles for proton 
momentum of $p_0$=24~GeV/c, data from~\cite{eich}.
\end{figure}

\textbf{New model for $\pi$-production from 5 GeV to 100 TeV.}
A new phenomenological model has been developed and introduced into 
\textsc{mars} as the default to describe pion production in high-energy 
proton-nucleus interactions~\cite{mokstr98}. 
Special attention is paid to low-momentum pions (p$<$2~GeV/c) from intermediate 
incident proton momenta (5$<p_0<$30~GeV/c). 
The following form is used for the double differential cross section
of the $pA \rightarrow \pi^{\pm}X$ reaction:
\begin{equation}
{d^2\sigma^{pA \to \pi^{\pm}X} \over dpd \Omega}=
R^{pA \to \pi^{\pm}X}(A,p_0,p,p_\perp){d^2\sigma^{pp \to
\pi^{\pm}X} \over dpd\Omega},                             
\end{equation}
where $p$ and $p_\perp$ are total and transverse momenta of $\pi^{\pm}$,
and $A$ is the atomic mass of the target nucleus.
The function $R^{pA \to \pi^{\pm}X}$, measurable with much higher precision
than the absolute yields, is almost independent of $p_\perp$  
and its dependence on $p_0$ and $p$ is much weaker than for the differential 
cross-section itself. 
Rather sophisticated algorithms have been developed 
to treat this function for pion production on nuclei in the forward 
($x_F>$0) and backward ($x_F<$0) hemispheres separately.
It is demonstrated in \cite{mokstr98} that model predictions are in a good
agreement with data in the entire kinematic region. 
Typical examples of comparison with data are shown in Figs.~2(b) and 3. 
Calculations with the \textsc{mars13(98)} code of the pion double differential 
spectra from a thick lead target at $p_0$=8~GeV/c agree reasonably well with 
data\,\cite{audus} in the {\em difficult} momentum region 0.5$<p<$5~GeV/c (see Fig.~3(b))
whereas \textsc{geant} seems to have a problem.

\begin{figure}[t!]
\begin{minipage}[t!]{.45\linewidth} 
\psfig{figure=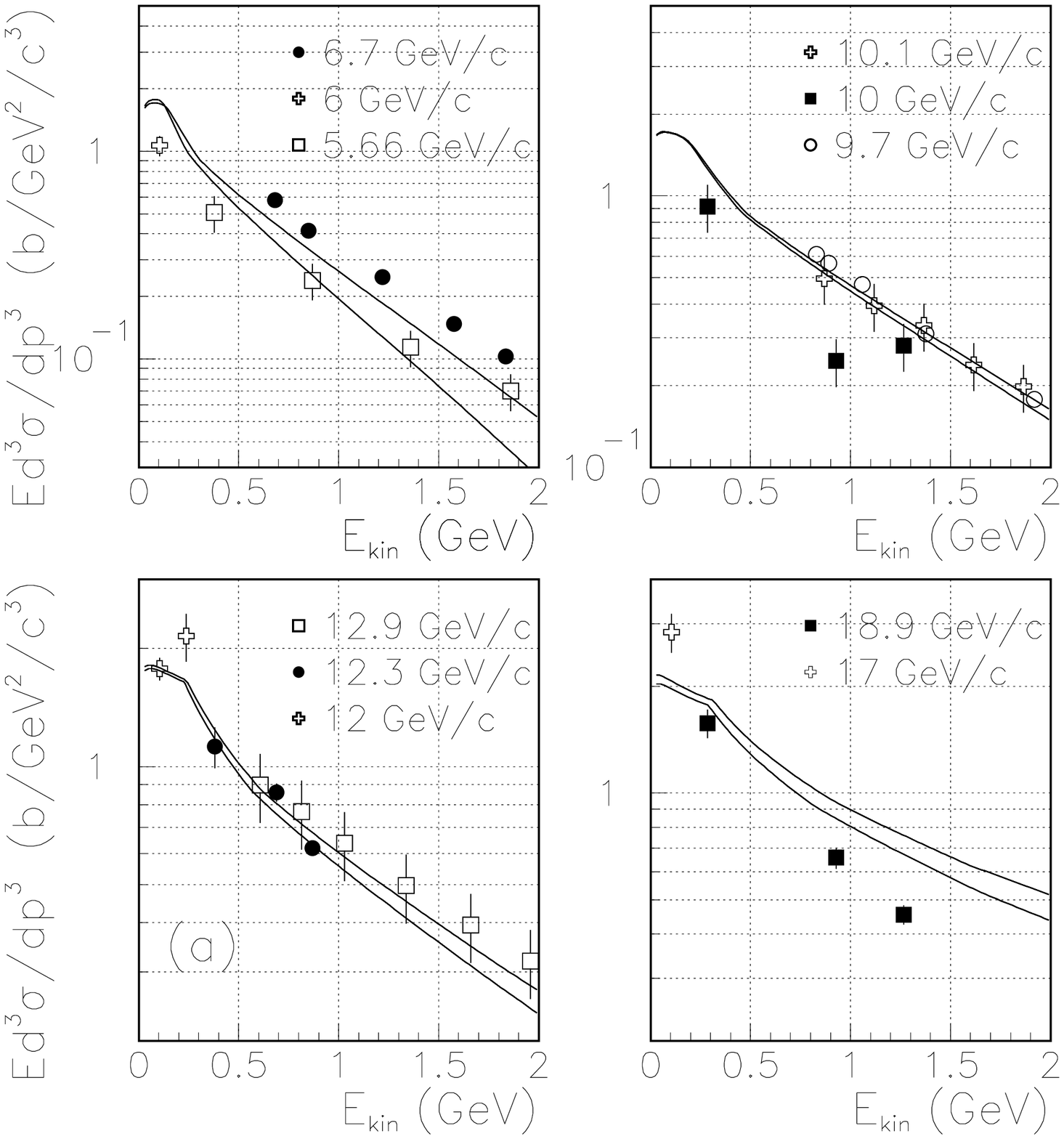,width=\linewidth,angle=0}
\end{minipage}\hfill
\begin{minipage}[t!]{.45\linewidth} 
\psfig{figure=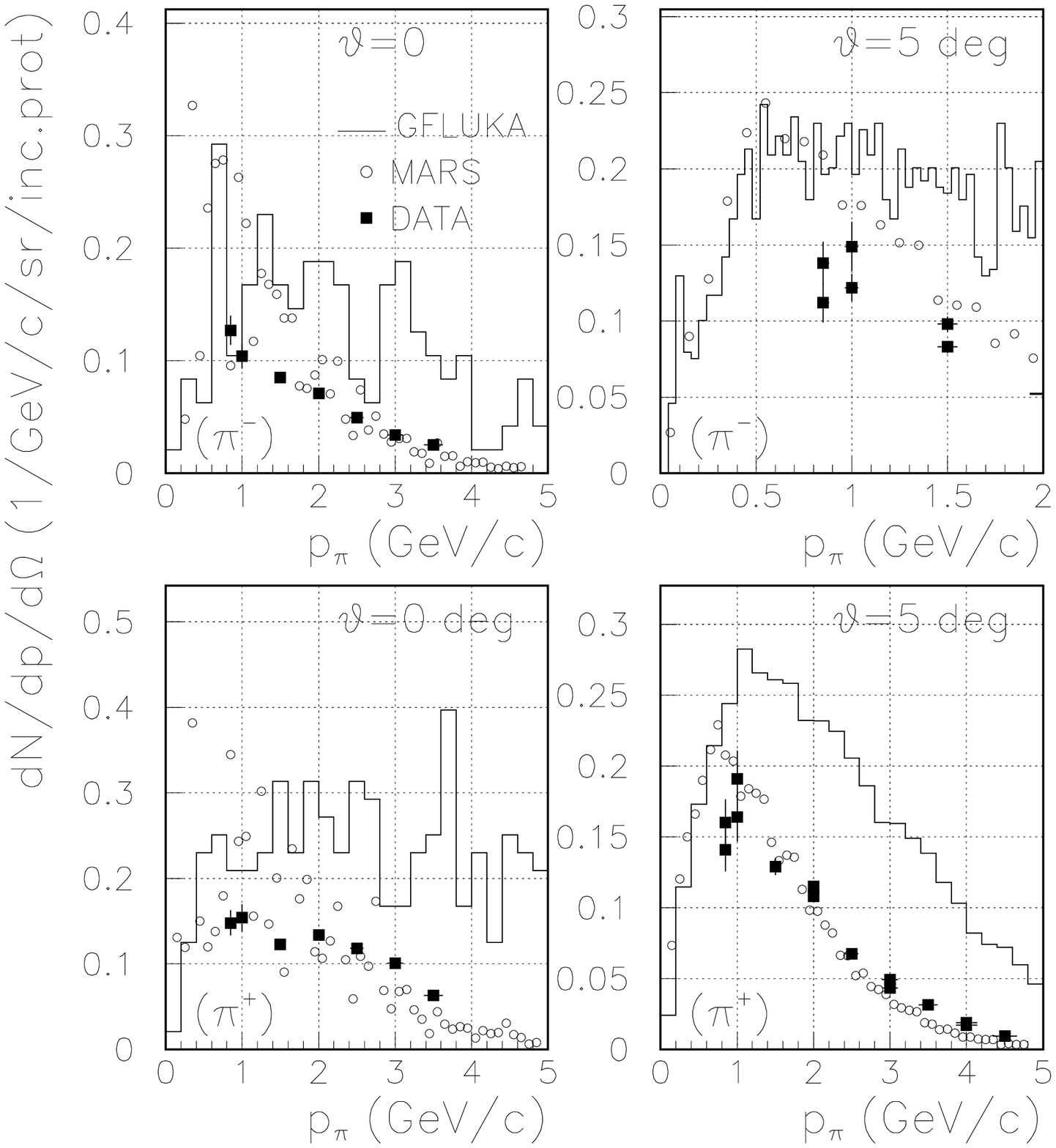,width=\linewidth,angle=0}
\end{minipage}

\vspace*{0.3cm}
Figure 3.
\textsc{mars} pion spectra {\em vs} data:
(a) $pCu \rightarrow \pi^-X$ ($\theta \approx$ 0), the two curves correspond to 
highest and lowest proton momentum indicated (see \cite{mokstr98} 
for references);
(b) $\pi^-$ (top) and $\pi^+$ (bottom) spectra from a 10-cm thick lead target 
at $p_0$=8~GeV compared with data\,\cite{audus} and
\textsc{geant} (\textsc{fluka} mode)~\cite{geant} prediction.
\end{figure}

\textbf{Cascade-exciton model code \textsc{cem95}.}
A version of the Cascade-Exciton Model of nuclear reactions~\cite{CEM}
as realized in the code \textsc{cem95}~\cite{CEM95} and
containing also several recent refinements~\cite{CEM98} is
now implemented as default for 1-10~MeV $<$ E $<$ 3-5~GeV.
The 1994 {\em International Code Comparison for Intermediate Energy Nuclear
Data} has shown that \textsc{cem95} adequately describes nuclear reactions at 
intermediate energies and has one of the best predictive powers for double 
differential cross sections of secondary particles as compared to other 
available models.
Besides that, it adds to \textsc{mars} reliable $\pi^-$-capture description 
(with a few modifications, e.g., radiative capture $\pi^{-}p \rightarrow  
n\gamma$), better description of photon induced reactions in the intermediate
energy range and of radionuclide production. 
To be usable in \textsc{mars}, the \textsc{cem95} code is converted 
into double precision along with some other necessary modifications.
Several examples of the \textsc{cem} predictions compared with experimental 
data and results of several other models are given in Figs.~4 and 5.
One can see that on the whole, the code reproduces quite well not only spectra
of secondary nucleons but also excitation functions for the spallation
yields, a 
much more {\em difficult} characteristic of nuclear reactions to be predicted 
by any theory, and is consistent with other well-known models~\cite{mash97}. 

\begin{figure}[b!]
\begin{minipage}[b!]{.33\linewidth} 
\psfig{figure=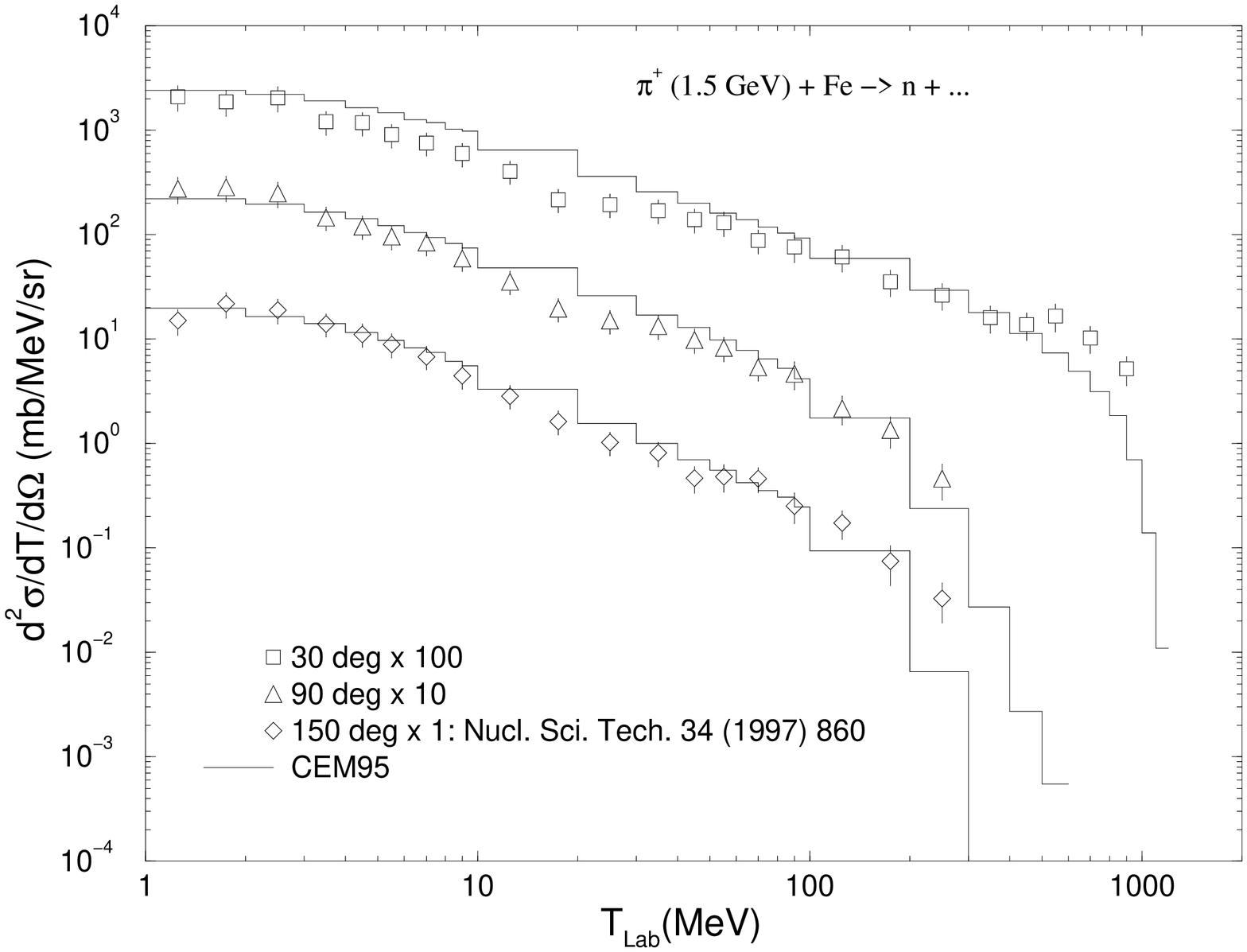,width=8cm,angle=-90}
\end{minipage}\hfill
\begin{minipage}[b!]{.33\linewidth} 
\psfig{figure=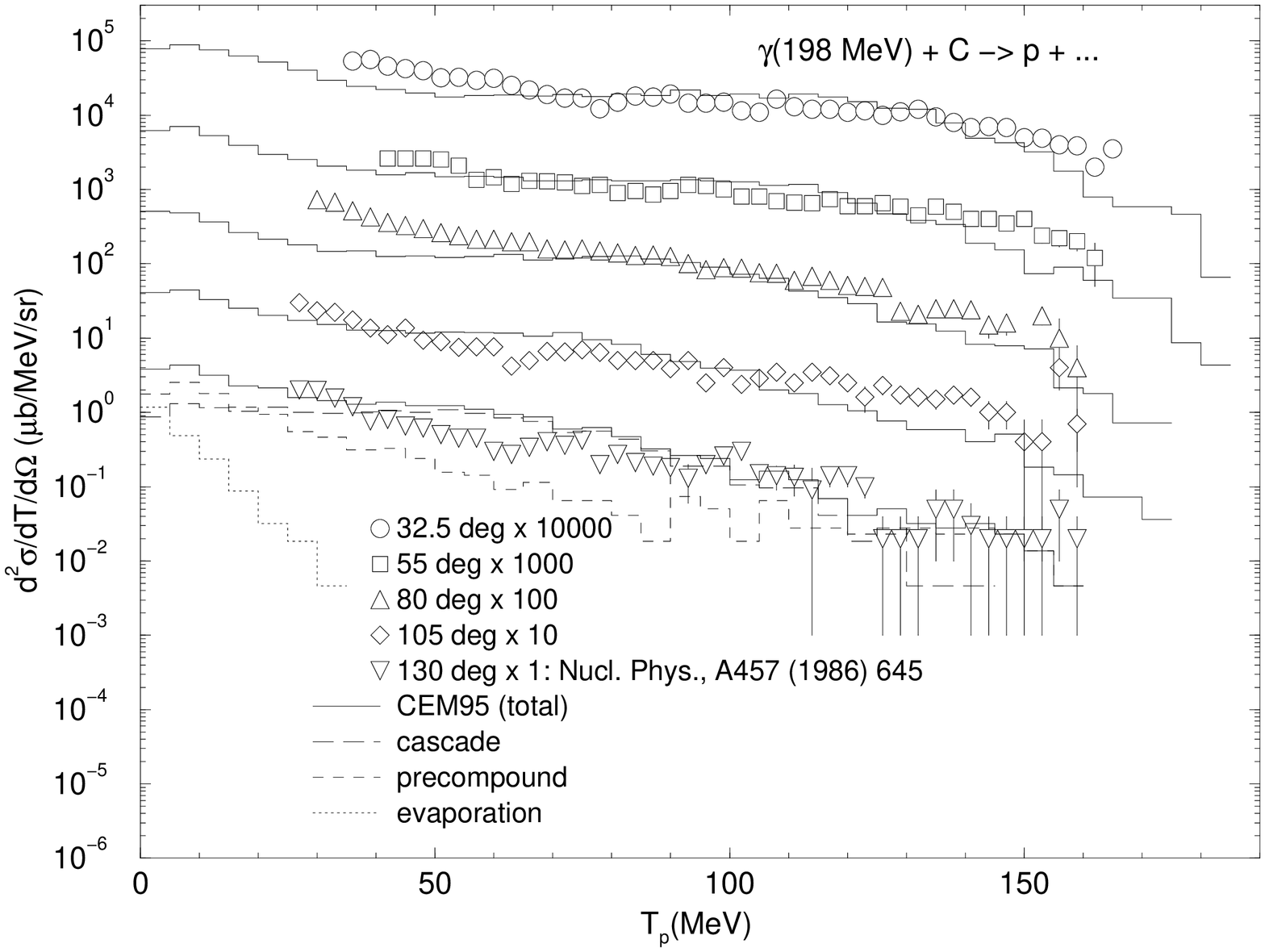,width=8cm,angle=-90}
\end{minipage}

\vspace*{0.3cm}
Figure 4.
Double differential nucleon spectra as calculated with \textsc{cem95}:
(a) $\pi^{+}Fe \rightarrow n+X$ for 1.5~GeV pions compared to data~\cite{nakamoto97};
(b) $\gamma C \rightarrow p+X$ for 198~MeV photons compared to data~\cite{gamC86}.
\end{figure}

\begin{figure}[t!]
\centerline{
\psfig{figure=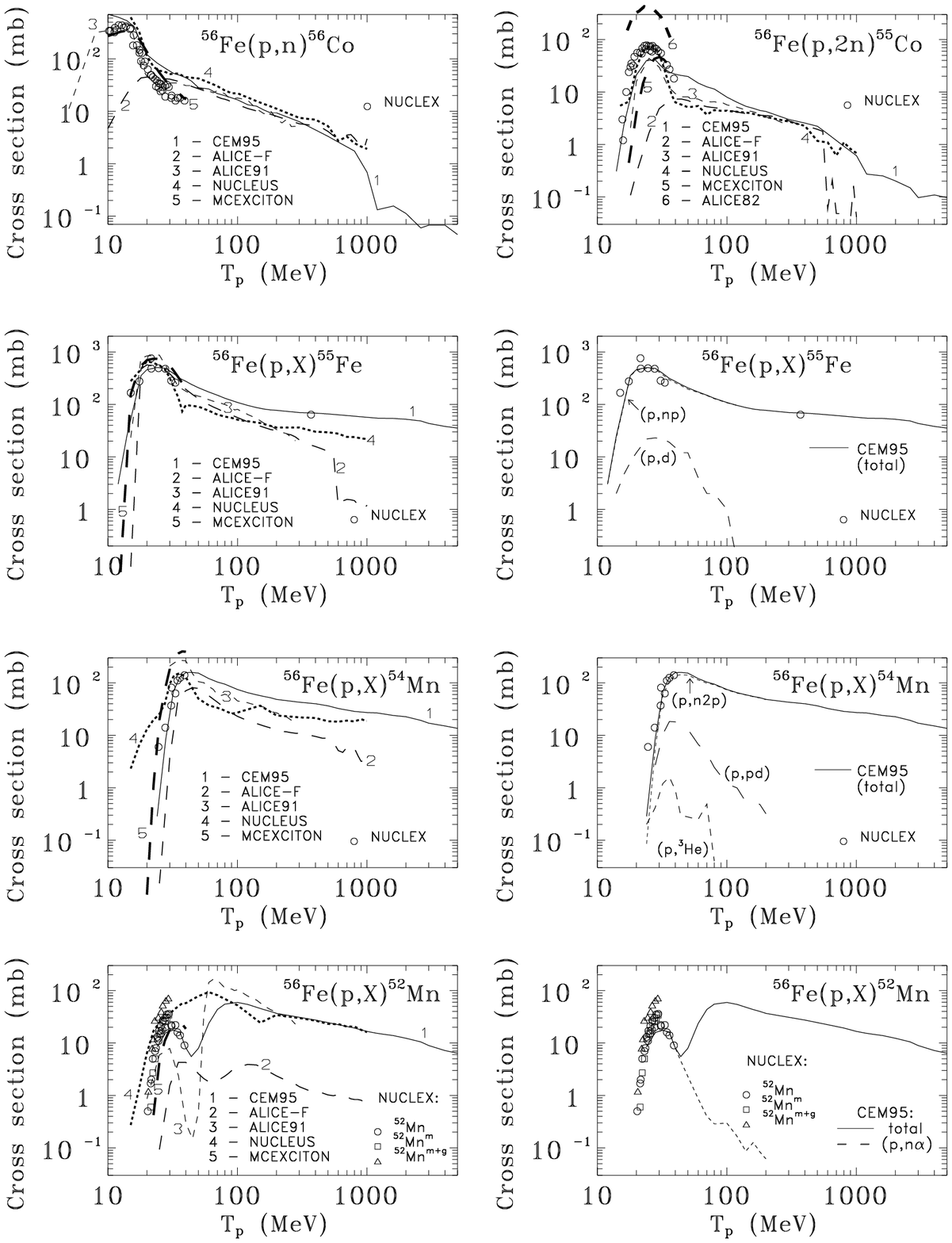,width=8cm,angle=0}
}

\vspace*{0.3cm}
Figure 5.
Excitation functions for the production of various radionuclides in
$p+^{56}Fe$ spallation reactions {\em vs} proton kinetic energy calculated
with \textsc{cem95} and 
compared with available experimental data and predictions of several
other models. For details and references see~\cite{mash97}.
\end{figure}

\textbf{\textsc{dpmjet} for primary interactions.}
The \textsc{dpmjet-2.3/2.4} code~\cite{dpmjet} is implemented
into \textsc{mars} to sample the initial $hN$, $hA$, $AA$ and $\nu A$
interaction for 5~GeV$<$E$<$100~TeV. This provides---at least partially---features
of a full exclusive event generation with all known particles in a final state.
The \textsc{dpmjet} code has been proven to be consistent with
collider and cosmic ray data in a multi-TeV energy region.
The two-component Dual Parton Model is used with multiple 
soft chains and multiple minijets at each elementary interaction. 
Within this model the high energy projectile undergoes a multiple scattering 
process as formulated in the Glauber approach. 
Particle production is realized by the fragmentation of colorless 
parton-parton chains constructed from the quark content of the interacting
hadrons. 
The code includes cascading of secondaries---suppressed by the formation time 
concept---within both target and projectile nucleus.
The excitation energies of the remaining target-- and projectile nuclei
are calculated and simulation of subsequent nuclear evaporation is included 
in the model. 
The coupling of these new features to the \textsc{mars} code is very 
\textsc{cpu}-time consuming and is used optionally only.

\textbf{Deuteron-nucleus collisions.}
Deuteron interactions have little in common with the general picture of
the interaction between complex nuclei because of the deuteron's relatively 
large size and small binding energy. 
Therefore a special model has been developed~\cite{deu98} and implemented 
into \textsc{mars}.
Deuteron-nucleus interactions are classified as elastic, dissociation, 
stripping, and full inelastic.
In elastic interactions the deuteron emerges intact in the final state while
the nucleus may be unchanged (coherent elastic) or have lost one nucleon
(incoherent).
Coherent elastic uses Glauber's treatment with some adjustments of the
parameters to fit experiment.
Incoherent elastic scattering assumes a differential cross section 
to be twice that of the proton---using the prescription of~\cite{bell}---and the nuclear 
parameters as for the coherent case.
This is then multiplied by a deuteron-- and a nucleon form factor as well
as a Pauli suppression factor which hinders low momentum transfers. 
Exchange of a long range virtual photon may result in Coulomb dissociation
whereby the deuteron splits into a proton and neutron.
This is calculated using a Weiszacker-Williams approach for virtual photon 
emission.
Dissociation may also result from (nuclear) elastic processes at relatively
high momentum transfers.
In stripping one nucleon undergoes an inelastic nuclear event while its
partner continues without interaction.
The total stripping probability is calculated based on the projected n--p 
separation as predicted by the deuteron wave function~\cite{hulth} and 
geometrical arguments.
Deuterons dissociate as in~\cite{serb} with full relativistic kinematics.
Interaction with the nucleus by one of the partners proceeds within
the standard \textsc{mars} scheme.
In full inelastic events both nucleons interact with the nucleus.
The stripping routine provides the angular deflection and momentum of each
nucleon after which both are allowed to interact as other \textsc{mars} 
nucleons.
As an example, a calculated $\pi, K$-meson yield out of a 3-cm radius gallium
target 36-cm long in a 7.5-cm radius solenoid (B=20~T) is presented in 
Fig.~6(a) for proton and deuteron beams of equal momentum per nucleon. 

\textbf{Neutrino-nucleus interactions.}
Neutrinos from a high energy muon collider may cause a radiation problem at 
large distances from the source~\cite{king96,nu97}. 
Extraterrestrial neutrinos have also been examined as radiation 
hazards~\cite{cos97}.
A special weighted neutrino interaction generator has been developed and 
incorporated into \textsc{mars}.
This model represents energy and angle of the particles---$e^{\pm}, \mu^{\pm}$, 
and hadrons---emanating from a simulated interaction.
These particles, and the showers initiated by them, are then further processed 
by the \textsc{mars} transport algorithms in the usual way.
The four types of neutrinos are distinguished throughout: $\nu_{\mu}, 
\overline{\nu}_{\mu}, \nu_{e}, \overline{\nu}_{e}$.
The model identifies the following types of neutrino interactions for
$\nu_{\mu}$ ($\overline{\nu}_{\mu}$) and similarly for $\nu_{e}, 
(\overline{\nu}_{e}$):
$\nu_{\mu}N \rightarrow \mu^{+}X,   \nu_{\mu}N \rightarrow \nu_{\mu}X,
\nu_{\mu}p  \rightarrow \mu^{+}n,   \nu_{\mu}p \rightarrow \nu_{\mu}p,
\nu_{\mu}n  \rightarrow \nu_{\mu}n, \nu_{\mu}e^{-} \rightarrow \nu_{\mu}e^{-},
\nu_{\mu}e^{-} \rightarrow \nu_{e}\mu^{-}, 
\nu_{\mu}A \rightarrow \nu_{\mu}A$. The formulas for these processes
as well as results of Monte Carlo simulations in general and in application
to muon colliders in particular are described in~\cite{nu98}.
Fig.~6(b) shows calculated dose in a 30-cm slab tissue-equivalent phantom
irradiated with neutrinos of 100~MeV$<E_{\nu}<$10~TeV energy.

\begin{figure}[t!]
\begin{minipage}[t!]{.48\linewidth} 
\psfig{figure=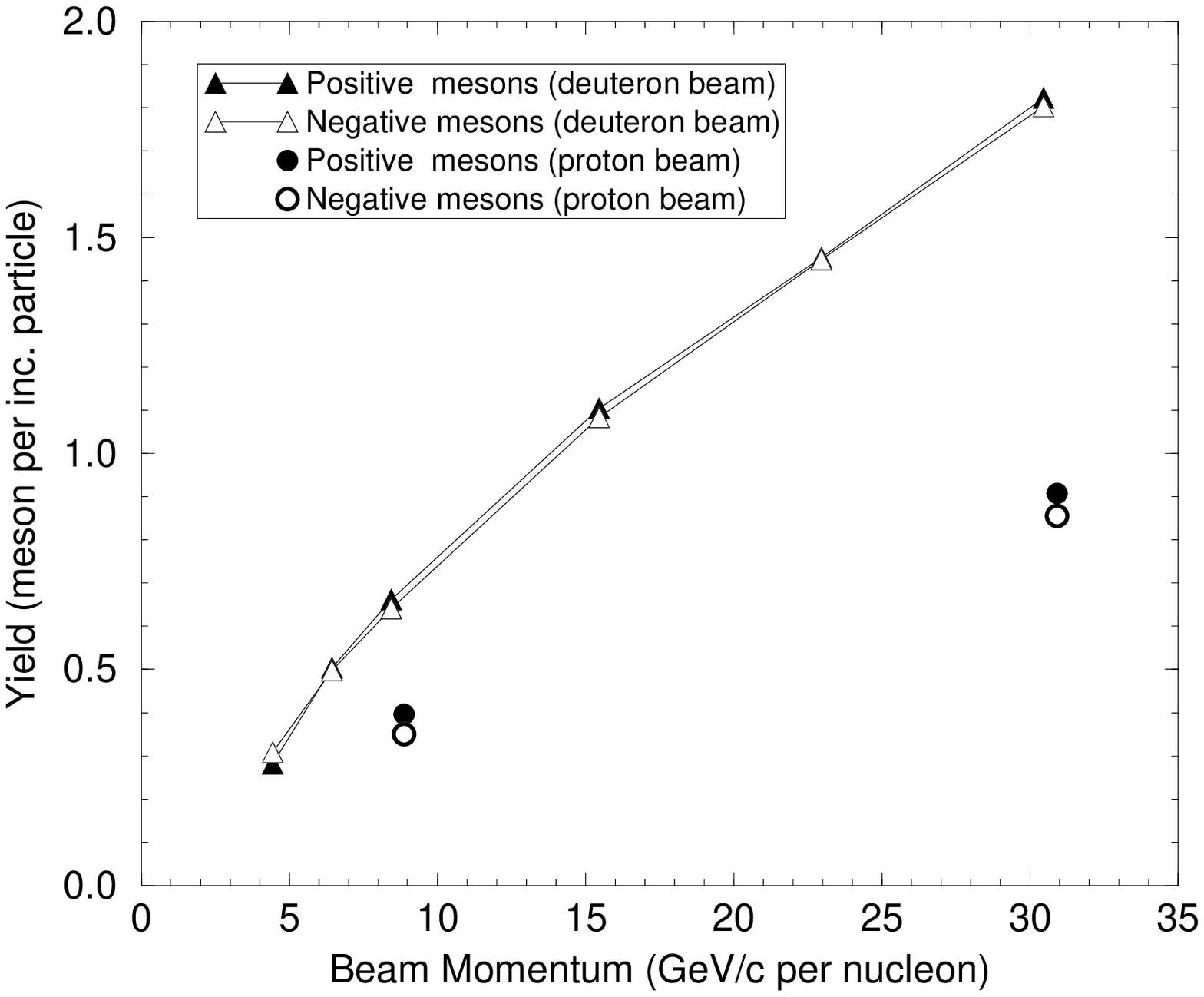,width=\linewidth,angle=0}
\end{minipage}\hfill
\begin{minipage}[t!]{.48\linewidth} 
\psfig{figure=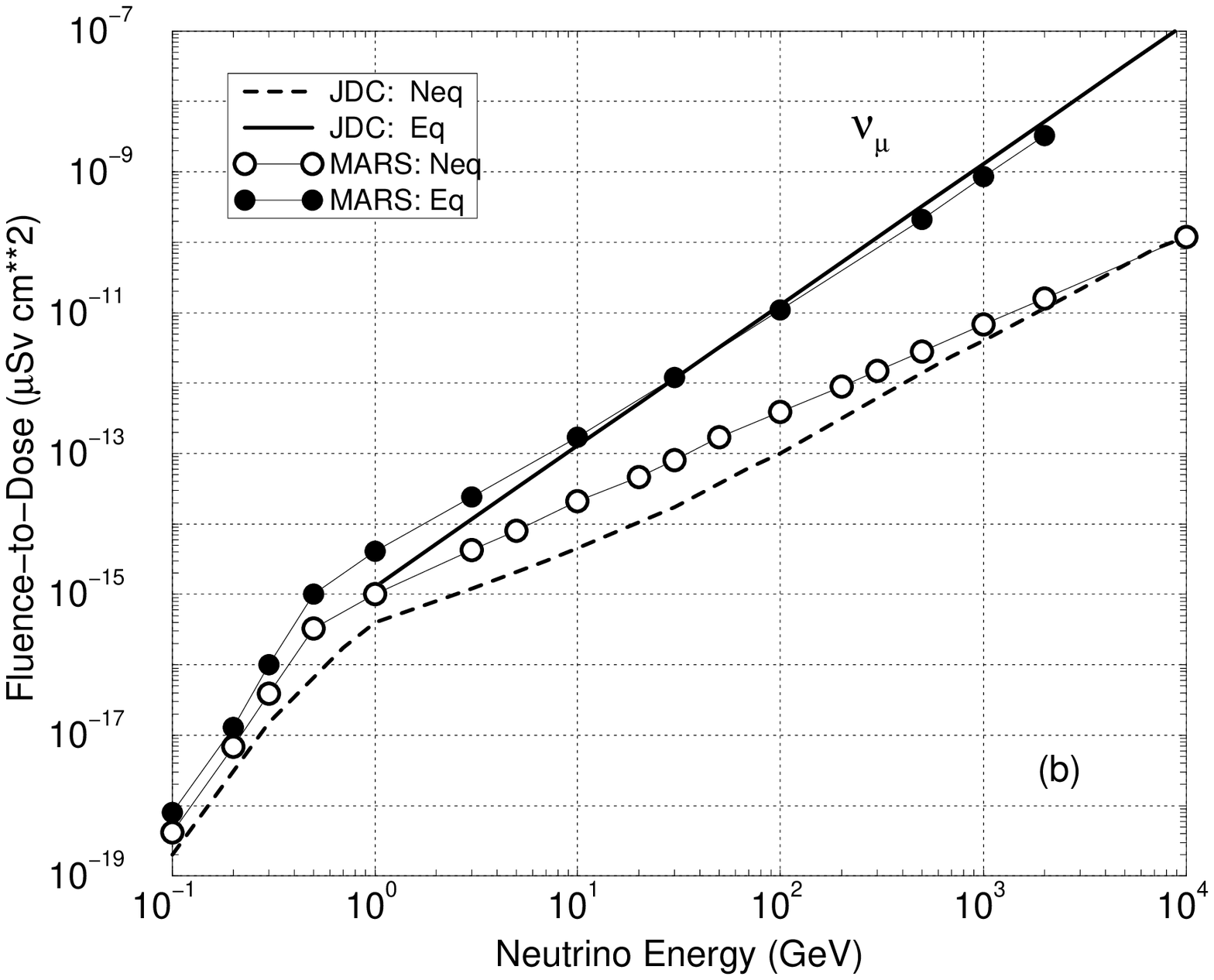,width=\linewidth,angle=0}
\end{minipage}

\vspace*{0.3cm}
Figure 6.
(a) $\pi, K$-meson yield from a gallium target for proton and 
deuteron beams calculated by {\sc mars};
(b) Dose equivalent in a bare (Neq) tissue phantom and in one embedded into 
an infinite medium (Eq) per unit neutrino fluence as calculated with 
\textsc{mars} and estimated in~\cite{cos97}.
\end{figure}

\section{Stopped Hadrons and Muons}

A very careful treatment is done in \textsc{mars} of processes near and below 
the Coulomb barrier in hadron and muon transport (ionization absorption $vs$ nuclear 
interaction $vs$ decay).

\textbf{Pions.}
A stopping $\pi^{+}$ decays into $\mu^{+}$ of 4.1~MeV plus a neutrino while
a $\pi^{-}$ attaches to a nucleus (via the modified Fermi-Teller law).
While cascading down the atomic energy levels, the pion is captured from a 
high orbit thus emitting only a few low energy photons which are neglected here.
The hadronic interaction of the stopped $\pi^{-}$ is treated using the 
Cascade-Exciton Model~\cite{CEM95} with a few modifications.
When hydrogen is the target it is assumed there is a 60\% probability to 
for charge exchange ($\pi^{-}p \rightarrow \pi^{0}n$) whereupon the $\pi^{0}$ 
decays into two photons of 68.9~MeV each and the neutron acquires a small 
(0.4~MeV) kinetic energy.
The remaining 40\% of stopped $\pi^{-}$ in hydrogen interact via
radiative capture: $\pi^{-}p \rightarrow  n\gamma$.
Here the photon acquires 129.4~MeV and the neutron 8.9~MeV kinetic energy.
Other nuclides have a much smaller probability for radiative capture (1--2\%
which is taken into account in competition with \textsc{cem95}).
The photon energy is chosen from an empirical fit to experiment while
the remainder is deposited as excitation energy.

\textbf{Muons.}
A stopping $\mu^{+}$ always decays into $e\nu\overline{\nu}$ while a 
$\mu^{-}$ attaches itself to a nucleus.
When a $\mu^{-}$ stops in a compound or mixture one first decides to 
which nucleus the $\mu^{-}$ attaches (modified Fermi-Teller law).
Following attachment the muon may still decay as decided by comparing 
capture and decay lifetimes of which the latter is favored for light
nuclei (Z$\leq$11).
A captured $\mu^{-}$ then cascades down to the ground state of the muonic 
atom emitting photons along with some Auger electrons, all of which is 
simulated using approximate fits to the atomic energy levels.
In hydrogen muon capture always produces a 5.1~MeV neutron via inverse
$\beta$-decay.
In complex nuclei the giant dipole resonance plays a role and results in 
an `evaporation'-type neutron spectrum with one or more resonances 
superimposed.
This is illustrated in Fig.~7(a) which shows the neutron spectrum resulting
from $\mu^{-}$ capture on oxygen.
In addition smaller numbers of evaporation-type charged particles and photons
may be  emitted. 
Calculated with the above algorithms longitudinal dose distributions in
a slab tissue-equivalent phantom are shown in Fig.~7(b) at the axis of
150~MeV proton and 75~MeV pion, muon and neutron beams striking the phantom.

\begin{figure}[b!]
\begin{minipage}[b!]{.48\linewidth} 
\psfig{figure=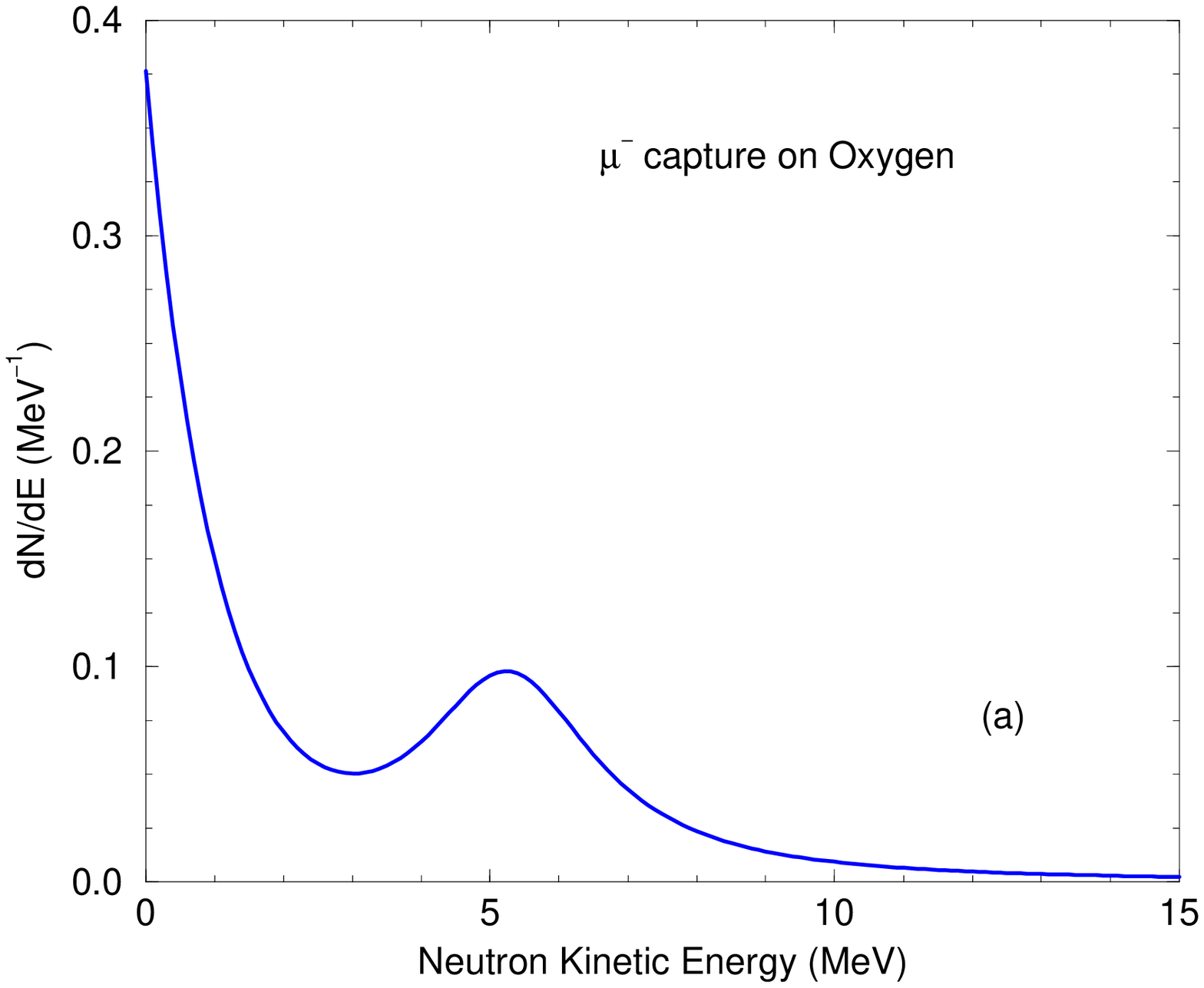,width=\linewidth,angle=0}
\end{minipage}\hfill
\begin{minipage}[b!]{.48\linewidth} 
\psfig{figure=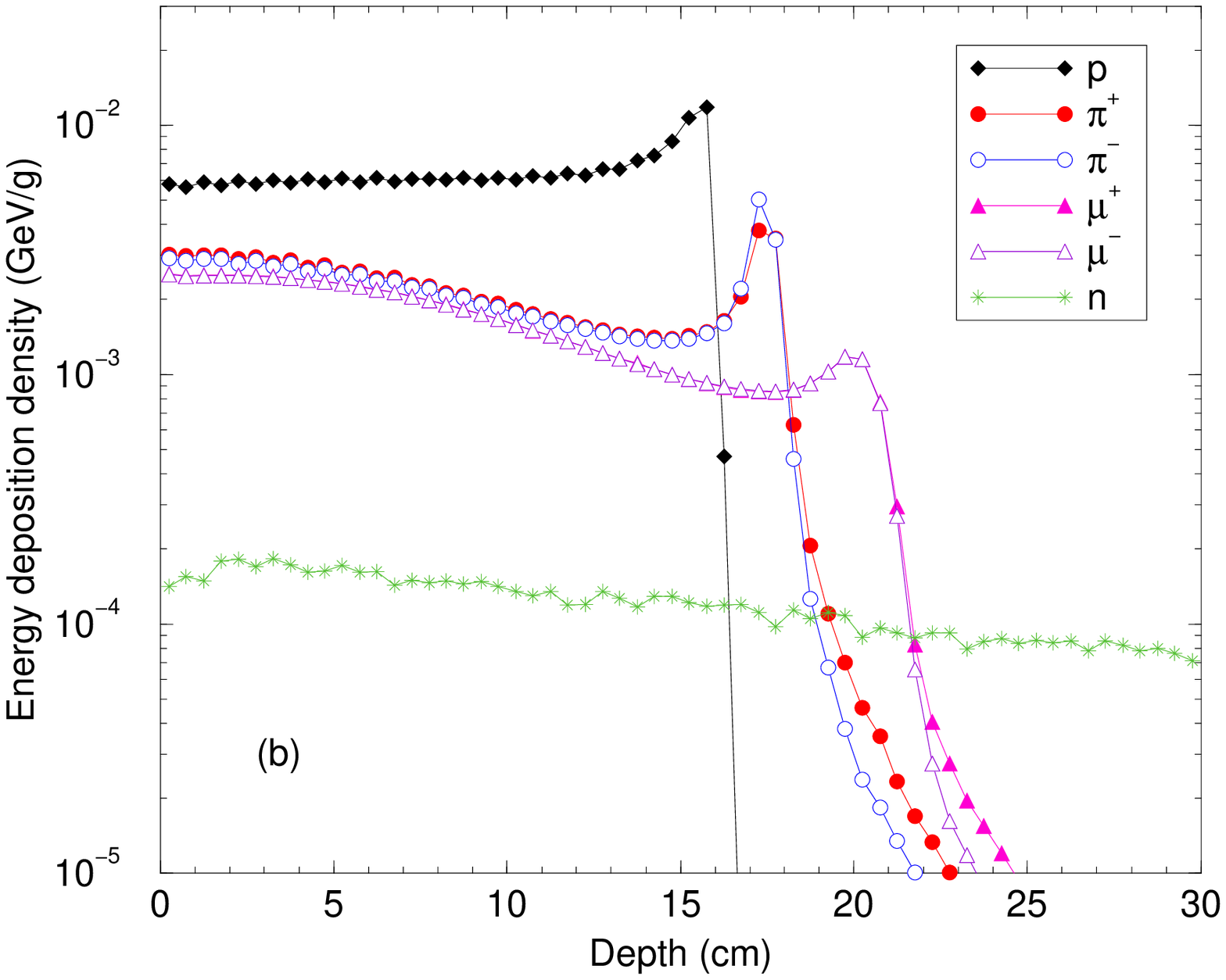,width=\linewidth,angle=0}
\end{minipage}

\vspace*{0.3cm}
Figure 7.
(a) Neutron spectrum generated in a $\mu^-$ capture on oxygen atom;
(b) Axial absorbed dose in a tissue-equivalent phantom for a 150~MeV ($p$)
and 75~MeV ($\pi^{\pm}, \mu^{\pm}, n$) 1$\times$1~cm beams.
\end{figure}

\newpage
\textbf{Antiprotons.}
Stopped $\overline{p}$ attach to nuclei in the same way as $\pi^{-}$
or $\mu^{-}$.
Annihilation at rest is assumed to produce only pions, neglecting some of
the rarer modes involving strange particles.
Charges of produced pions are slightly skewed towards $\pi^{-}$ in view of the
`brought in' negative charge.
Pion momenta are chosen from an inclusive distribution loosely based on 
experiment.
The energy weighted distribution is normalized to twice the nucleon mass
which predicts a multiplicity of 4.3---close to observation.
In a complex nucleus the annihilation is treated as though it occurs on free 
nucleon except that each pion produced by the annihilation process is given 
a 50\% probability to interact within the nucleus.
This shortcut attempts to include---at least qualitatively---participation by
the constituents nucleons.

For antiprotons in flight the annihilation cross section results in a larger
cross section for $\overline{p}A$ vis-a-vis $pA$, especially for light nuclei
at lower energies.
Total cross sections for both $\overline{p}A$ and $pA$ are estimated on the 
basis of simple geometrical considerations and $\overline{p}p$, 
$\overline{p}n$ and $pp$,$pn$ cross sections.
The ratio $\sigma_{\overline{p}A}/\sigma_{pA}$ is then applied 
to the more accurate $\sigma_{pA}$ used in \textsc{mars}.
Annihilation in flight uses the same inclusive pion
distribution as at rest in the $\overline{p}$-nucleon rest frame after which
the pions are Lorentz transformed back to the lab.
Above about 0.1~GeV/c a small $\overline{p}p \rightarrow \overline{n}n$
component is included.
For both mechanisms {\em nuclear} target effects are again approximated by 
allowing emerging particles to interact in the same nucleus or escape each 
with one half probability.
There is also added a third component in which the $\overline{p}$ or
$\overline{n}$ interact only quasi-elastically with the nucleons.
These are simulated using conventional \textsc{mars} algorithms exactly as for protons 
except that the fastest nucleon emerging (leading particle) from the collision is
identified as its antiparticle.

\section{Electromagnetic Processes.}

\hspace{6mm}
\textbf{Unified treatment for hadrons and muons.}
In muon and charged hadron electromagnetic processes, i.e., ionization and 
radiative collisions, with energy transfer $\varepsilon$ greater than a 
cutoff $\varepsilon _c$
are considered as discrete events involving production of
$\delta$-electrons, e$^{+}$e$^{-}$-pairs, and bremsstrahlung~\cite{mokstr96}.
The secondaries with $E > \varepsilon _c$ are followed explicitly.
Energy losses with $ \varepsilon < \varepsilon _c$ (so-called restricted
losses) are considered as continuous. 
The restricted loss distribution is described by Vavilov's function
with redefined parameters
 \begin{eqnarray}
 \xi = B s, &&
  B = 0.1536 {Z \over A \beta^2} \nonumber\\
  \kappa_n = \xi/\varepsilon _c, &&
  \beta^2_n = \beta^2 \varepsilon _c/\varepsilon _{\rm max}
\end{eqnarray}
where  $Z$ and $A$ are the atomic and mass numbers of the absorber,
$\beta c$ is the particle velocity, $\varepsilon _{\rm max}$ is 
the maximum energy transferred in a single collision, where
the constant $B$ has dimensions $MeV \cdot g^{-1} \cdot cm^{2}$ so that,
with the path-length $s$ expressed in g/cm$^2$, $\xi$ is in MeV. 
However, when $\kappa_n > 10$, the distribution
becomes approximately Gaussian with mean
\begin{equation}
\bar\Delta_r=\alpha(\varepsilon<\varepsilon_c) \cdot s
\end{equation}
and variance
\begin{equation}
\sigma^2_r=\frac{\xi^2}{\kappa_n}(1-\frac{\beta^2_n}{2})
\end{equation}
where $\alpha(\varepsilon<\varepsilon_c)$ is the mean restricted energy loss
per unit length. 
The Gaussian approximation drastically simplifies the simulation.
To simulate $\delta$-electron production at any step, one calculates
$\varepsilon _{\delta}= \min (\varepsilon _G,\varepsilon_c)$ where
$\varepsilon_G = \xi /10$. 
The restricted energy loss with $E < \varepsilon _{\delta}$ is then sampled 
from a Gaussian.
The number of $\delta$-electrons with $E > \varepsilon _{\delta}$ is 
simulated using a Poisson distribution.
The energy of each $\delta$-electron---if any---is sampled from Bhabha's 
formula. 
Total energy loss of a particle is the sum of the $\delta$-electron energies 
and the restricted energy loss.
Calculated energy spectrum and angular distributions of 100 and 300~MeV muons
after lithium absorbers are shown in Fig.~8 in comparison with other 
calculations.

\textbf{Bremsstrahlung.} For muon bremsstrahlung, a precise but complex 
formalism~\cite{andreev78} used previously as described 
in~\cite{mokstr96} has been replaced with a new simplified 
algorithm~\cite{andreev97}, which practically coincides with the exact formula.
Total cross section and $dE/dx$ calculated by the two methods agree within 1\%.

\begin{figure}[t!]
\begin{minipage}[b!]{.45\linewidth} 
\psfig{figure=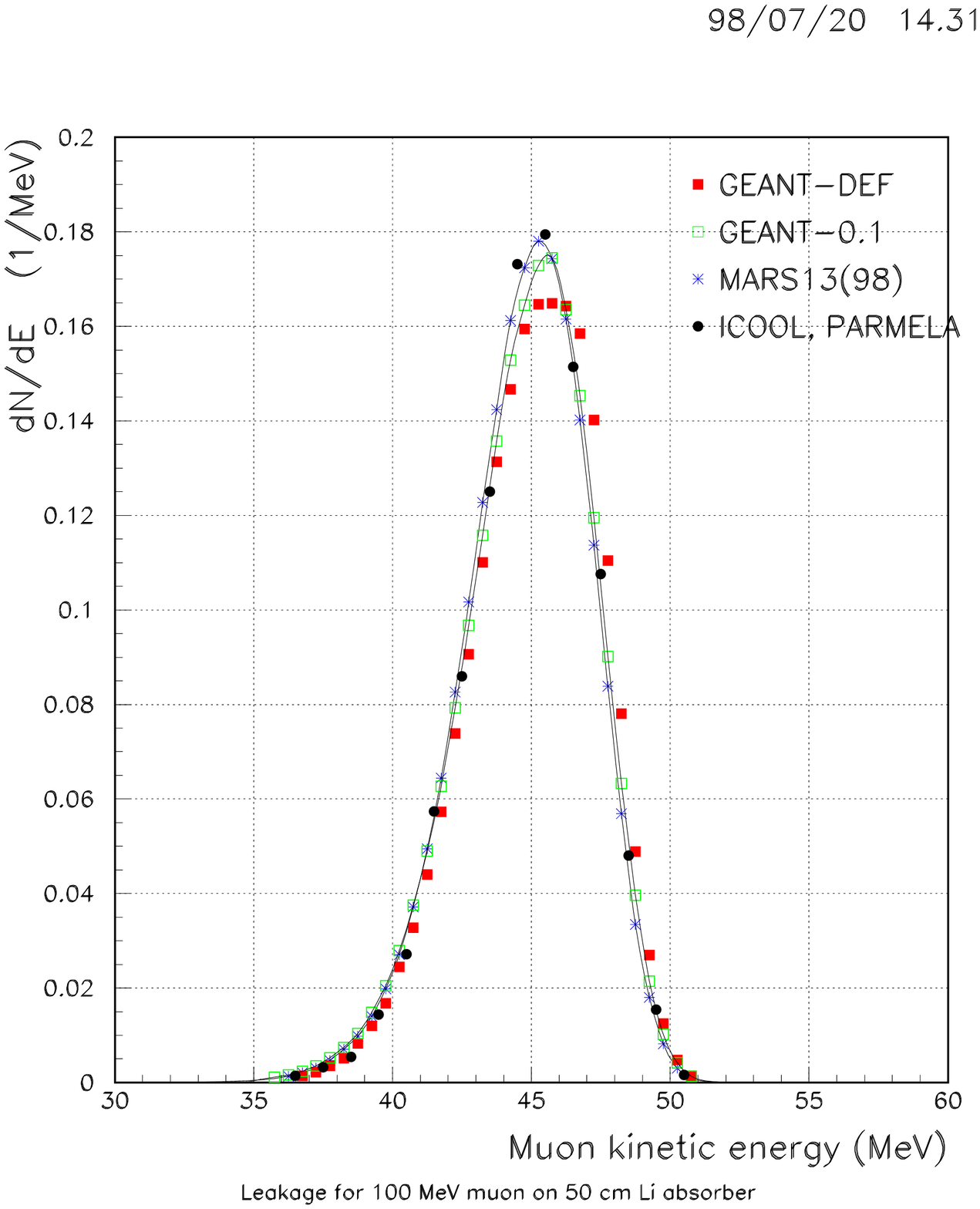,width=\linewidth,angle=0}
\end{minipage}\hfill
\begin{minipage}[b!]{.46\linewidth} 
\psfig{figure=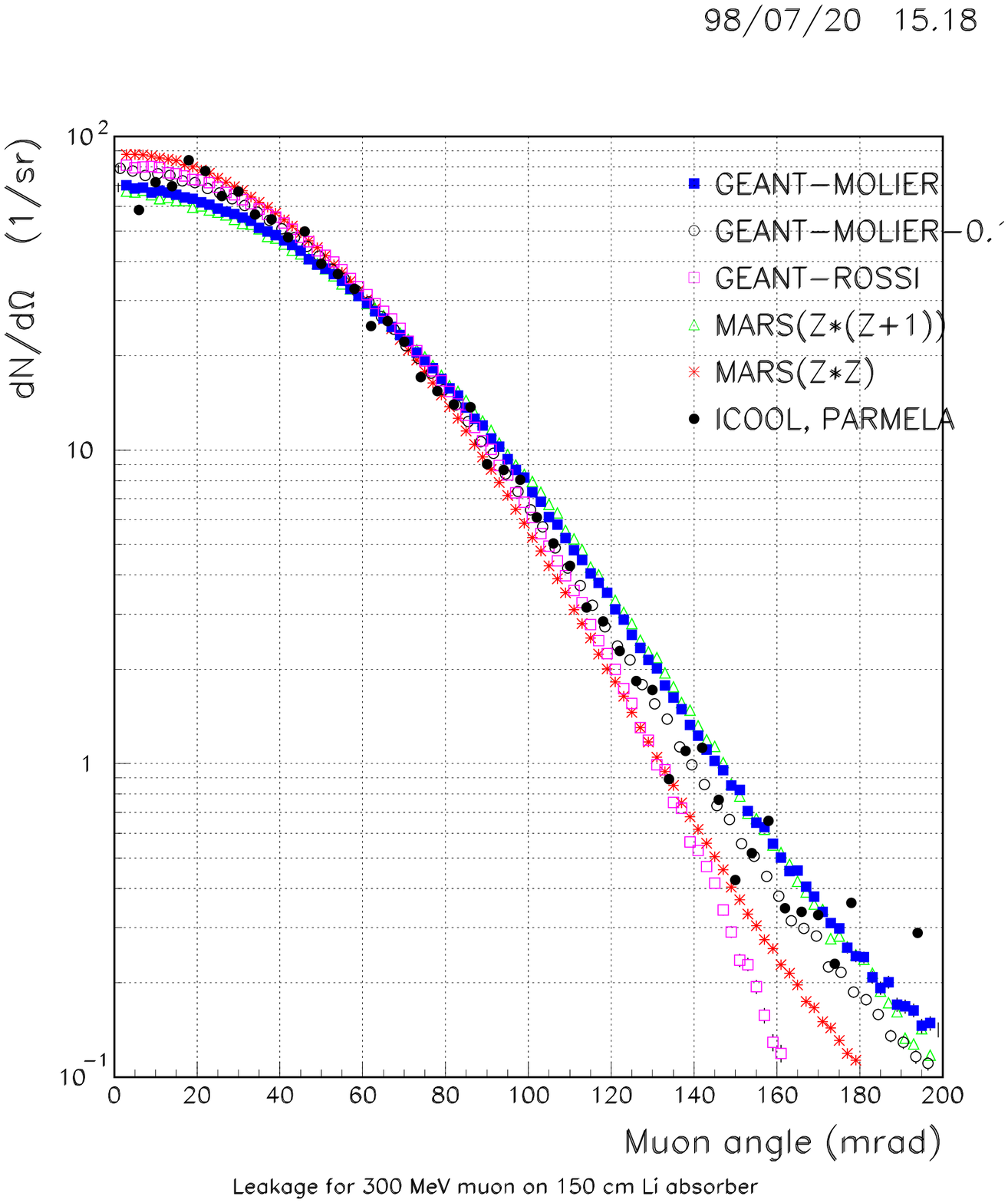,width=\linewidth,angle=0}
\end{minipage}

\vspace*{0.3cm}
Figure 8.
(a) Muon energy spectrum after a 50-cm lithium absorber for 100~MeV
incident muons as calculated with \textsc{mars}, \textsc{geant}~\cite{geant} (in two modes)
and \textsc{icool}~\cite{icool};
(b) Muon angular distribution after a 150-cm lithium absorber for 300~MeV
incident muons as calculated with \textsc{mars}, \textsc{geant} and \textsc{icool} 
in a few modes.
\end{figure}

\section{Other Enhancements}

\hspace{6mm}
\textbf{Precise treatment of mixtures and compounds.}
Precise treatment of individual elements in mixtures and compounds defined
through the weight-- or atomic fractions is done for all the electromagnetic 
and nuclear elastic and inelastic processes. 
Homogenization (averaging) thus becomes obsolete and is to be strongly 
discouraged.
Atomic masses for 100 elements of the periodic table~\cite{RPP98} and 
mean ionization energies and Sternheimer parameters~\cite{ICRU37,ADNDT84} 
are used by the code. 
All needed nuclear and electromagnetic cross sections and other parameters 
are calculated at the initialization stage for the specified composition of 
all materials present in the problem with a sample printout provided.
Corresponding array sizes are adjusted according to energy. 
Up to 50 composite materials may be present in a given run.

\textbf{Automatic \textsc{mars-mcnp} interface.} The code now includes an automatic
interface of \textsc{mars} materials to the \textsc{mcnp} code~\cite{MCNP}
for transport of low-energy neutrons and photons.

\textbf{List~of~particles.}~\textsc{mars}~now~includes
$p, n, \pi^+, \pi^-, K^+, K^-, \mu^+, \mu^-, \gamma, e^-, e^+, \overline p,
\pi^0, d, t, He^3, He^4, \nu$. 
In this version---for transport and output---$\overline n$ are included along
with the neutrons, heavy fragments deposit their energy locally and
short-lived hadrons and resonances are assumed to decay instantly into
the appropriate particles which are then followed by the standard \textsc{mars}
algorithms.

\begin{sloppypar}
\textbf{Geometry, visualization and histograming.} The Object-Oriented 
geometry engine and visualization module~\cite{mars97,Geom} is further 
developed. 
Histograming and visualization capabilities of results, including a 
geometry-independent mesh-based scoring tool, are further extended. 
An interface to the \textsc{ansys} code~\cite{ANSYS} for thermal and stress 
analyses~\cite{mars97} for use in certain applications is slightly improved.
\end{sloppypar}

\textbf{Variance reduction.} Algorithms for splitting and Russian roulette at 
$hA$ vertices and in particle transport are further improved. 
For `deep penetration' problems in complex highly non-uniform geometries, 
algorithms for scoring probabilities rather than real particle 
crossings or interactions now take into account all possible processes for 
both stable and unstable particles and charged as well as neutral hadrons.

\textbf{Recent benchmarking}
Fig.~9(a) shows results of a recent SATIF-4 benchmarking for attenuation of 
dose equivalent (E$>$20~MeV neutrons only) in a thick concrete absorber for 
parallel almost mono-energetic neutron beams~\cite{hideo98} of 100
and 400~MeV.
Results calculated with {\sc hetc}, {\sc anisn}, {\sc nmtc}, {\sc mcnp},
and {\sc mars} agree quite well over six decades at 400~MeV.
At 100~MeV wider discrepancies are observed.
Another \mbox{SATIF-4} benchmarking has been performed for a 30-cm thick
tissue-equivalent phantom irradiated with parallel mono-energetic neutron 
beams of 0.1 to 10~GeV energy~\cite{yoshizawa}. 
Fig.~9b shows \textsc{mars} results rather close to those of 
\textsc{hetc-3step}.

\begin{figure}[t!]
\begin{minipage}[b!]{.48\linewidth} 
\psfig{figure=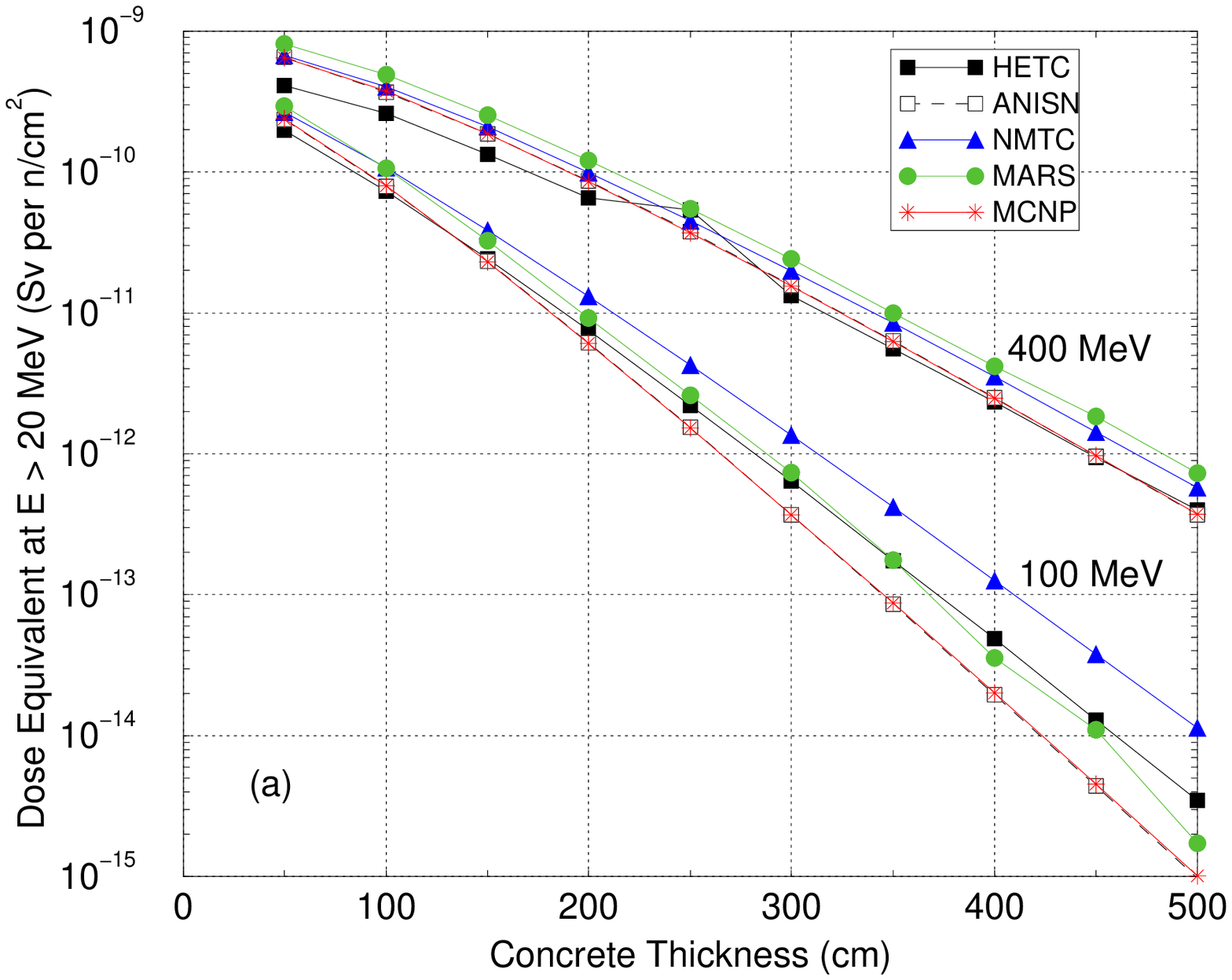,width=\linewidth,angle=0}
\end{minipage}\hfill
\begin{minipage}[b!]{.48\linewidth} 
\psfig{figure=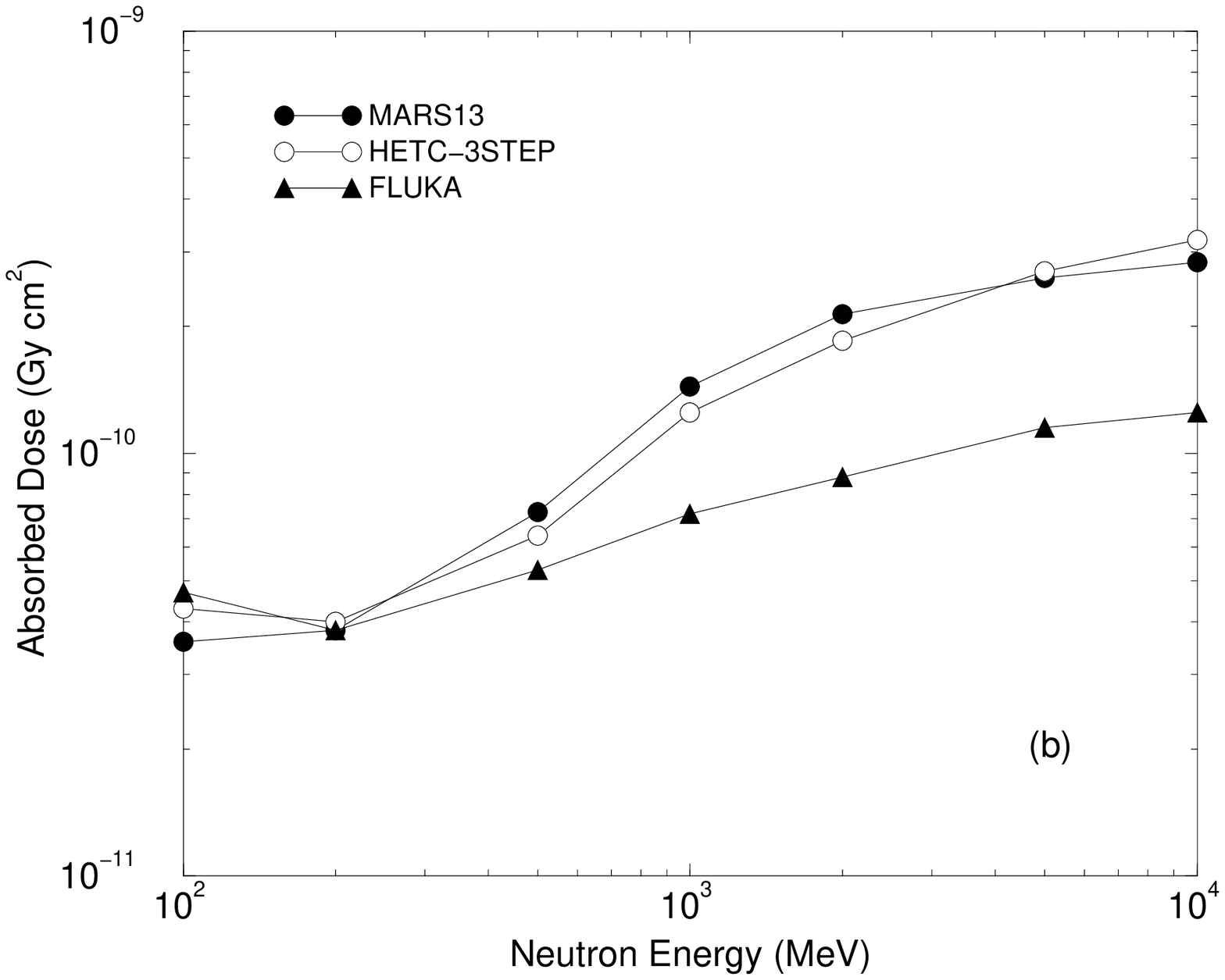,width=\linewidth,angle=0}
\end{minipage}

\vspace*{0.3cm}
Figure 9.
(a) Dose equivalent due to neutrons (E$>$20~MeV) in a thick concrete
absorber irradiated with 100 and 400~MeV neutron beams as calculated with {\sc hetc},
{\sc anisn}, {\sc nmtc}, {\sc mars} and {\sc mcnp} codes;
(b) Absorbed dose per unit neutron fluence at 1-cm depth of a slab tissue-equivalent 
phantom {\em vs} neutron energy calculated with the \textsc{mars},
\textsc{hetc-3step} and \textsc{fluka} codes.
\end{figure}

\section{Conclusions}

The \textsc{mars} code physics and scoring possibilities have been 
substantially extended.
The code's reliability is confirmed by several contributions to the SATIF/SARE
meetings as well as in several more recent applications 
where \textsc{mars} predictions show very good agreement with data.
The official \textsc{mars} site on the World Wide Web is 
{\em http://www-ap.fnal.gov/MARS/} which contains
information about the code, its users, and its uses.
At this site one can also register as a user and download the code for 
various platforms.

\end{document}